\documentclass[a4paper,12pt]{article}
\usepackage{setspace}
\usepackage{authblk}

\begin{document}

\title{%
Crystal growth and anisotropic transport properties of high-$T_c$ superconductors Bi$_{2}$Sr$_{2}$Ca$_{n-1}$Cu$_{n}$O$_{2n+4+\delta}$ ($n$ = 2, 3)
}
\author[1]{Takao Watanabe \thanks{E-mail address: watataka@aecl.ntt.co.jp}\thanks{Present address: NTT Photonics Laboratories, 3-1, Morinosato Wakamiya, Atsugi-shi, Kanagawa 243-0198, Japan}}
\author[2]{Takenori Fujii \thanks{ Present address: Cryogenic Center, University of Tokyo, 2-11-16, Yayoi, Bunkyo-ku, Tokyo 113-0032, Japan}}
\author[1,2]{Azusa Matsuda}
\affil[1]{NTT Basic Research Laboratories, 3-1, Morinosato Wakamiya, Atsugi-shi, Kanagawa 243-0198, Japan}
\affil[2]{Department of Applied Physics, Faculty of Science, Science University of Tokyo, 1-3 Kagurazaka, Shinjyuku-ku, Tokyo 162-8601, Japan}

\date{\today}

\maketitle
\begin{abstract}
Large high-quality single crystals of Bi$_{2}$Sr$_{2}$Ca$_{n-1}$Cu$_{n}$O$_{2n+4+\delta}$ ($n$ = 2, 3) were successfully grown using an improved traveling solvent floating zone (TSFZ) method, which features a slow growth rate and steep temperature gradient along the melting zone. By measuring anisotropic resistivities and susceptibilities of Bi$_{2}$Sr$_{2}$CaCu$_2$O$_{8+\delta}$, the characteristic pseudogap temperature $T^*$ was studied as a function of doping. The $T^*$ suggest that the pseudogap is not simply a precursor of high-$T_c$ superconductivity, but that the pseudogap and the superconducting gap compete with each other. The anisotropic resistivities of Bi$_{2}$Sr$_{2}$Ca$_2$Cu$_3$O$_{10+\delta}$ were also measured, revealing that the $T_c$ remains fixed in the overdoped region while anisotropy decreases continuously. This anomalous behavior will be discussed in terms of the inequivalent hole doping, which occurs between two inequivalent CuO$_2$ planes in the triple-layer system. 
\end{abstract}

\section{INTRODUCTION}
Ever since high-$T_c$ superconductivity was discovered more than 15 years ago \cite{bed}, intensive investigations have been carried out in attempts to reveal its mechanism. However, the mechanism of high-$T_c$ superconductivity has remained elusive because we have not been able to resolve some important issues in its physics. Resolving these issues has been difficult because of the lack of large high-quality single crystals, which are needed so that we can examine the anisotropic properties more precisely. In a case that the crystal is a new one, like Bi$_{2}$Sr$_{2}$Ca$_2$Cu$_3$O$_{10+\delta}$ (Bi-2223) here, the benefits of the successful crystal growth is especially large.

One unresolved issue in the physics of high-$T_c$ superconductivity is the pseudogap that depletes the electronic density of states (DOS) around the Fermi level from characteristic temperatures ($T^*$) far above the superconducting transition temperature ($T_c$). Since the pseudogap evolution is commonly observed in most high-$T_c$ superconductors \cite{tim}, it is generally believed that the pseudogap is closely related to the pairing mechanism of high-$T_c$ superconductivity. Therefore, the pseudogap phenomenon has been attracting much attension and a huge amount of experimental data has been accumulated. However, up to now, as shown in Fig. 1, there is still no consensus on when (at what temperatures) the pseudogap opens. One of the reasons for the lack of consensus is that there is probe dependence. Many probes such as angle resolved photoemission spectroscopy (ARPES) \cite{din,mar}, NMR \cite{war,ish} and in-plane resistivity $\rho_a$ \cite{tak} suggest that the $T^*$ is close to line "A". On the contrary, the uniform susceptibility $\chi$ of La$_{2-x}$Sr$_x$CuO$_4$ \cite{joh1} suggests that the $T^*$ is close to line "B". Another reason is that there is material dependence. For example, the $\chi$ of YBa$_2$Cu$_3$O$_{6+\delta}$ \cite{lor1} suggests that the $T^*$ is close to line "A". Clarifying whether the pseudogap phase boundary $T^*$ is "A" or "B" is crucially important because it would elucidate the relationship between the pseudogap and superconducting gap. The $T^*$ of boundary "A" implies that the pseudogap is a precursor of high-$T_c$ superconductivity, while that of the boundary "B" implies that the pseudogap and superconducting gap compete with each other. To find the generic pseudogap phase boundary, the Bi$_{2}$Sr$_{2}$CaCu$_2$O$_{8+\delta}$ system has been studied in detail using its large high-quality single crystals. 

Another unresolved issue is the physical origin of the correlation between the maximum $T_c$ and the number ($n$) of CuO$_2$ planes. Empirically, the maximum $T_c$ first increases with $n$ for $n\leq3$, then decreases \cite{kar}. Taking the Bi family, i.e. Bi$_{2}$Sr$_{2}$Ca$_{n-1}$Cu$_{n}$O$_{2n+4+\delta}$ ($n$ = 1-3), as an example, the maximum $T_c$ is approximately 34, 90, and 110 K for $n$ = 1 (Bi$_{2}$Sr$_{2}$CuO$_{6+\delta}$:Bi-2201), $n$ = 2 (Bi$_{2}$Sr$_{2}$CaCu$_2$O$_{8+\delta}$:Bi-2212), and $n$ = 3 (Bi$_{2}$Sr$_{2}$Ca$_2$Cu$_3$O$_{10+\delta}$:Bi-2223), respectively. Single crystals of triple-layered Bi-2223, however, had been very difficult to grow and large samples for physical property measurements were therefore unavailable for a long time. The difficulty is attributed to the narrowness, both in temperature and composition, of the liquidus line for Bi-2223. To overcome this serious problem, we have developed an improved traveling solvent floating zone (I-TSFZ) method.

The next section describes the concept of the I-TSFZ method and its successful application to Bi$_{2}$Sr$_{2}$Ca$_{n-1}$Cu$_{n}$O$_{2n+4+\delta}$ ($n$ = 2, 3) systems. Section 3 addresses, based on the systematic measurements of those crystals, several important issues on high-$T_c$ superconductivity: the pseudogap phase boundary, the origin of semiconducting out-of-plane transport $\rho_c$($T$), the origin of anomalous temperature/doping dependence of magnetic uniform susceptibility $\chi$, and the authenticity of charge confinement. In addition, for Bi-2223, anomalous doping dependence, in which $T_c$ does not decrease with increasing doping in the overdoped region, will be revealed. The reason for the anomalous behavior as well as the reason why $T_c$ is higher in triple-layered compounds will be discussed. Section 4 summarizes the paper.

\section{CRYSTAL GROWTH}
\subsection{FLUX METHOD vs. TSFZ METHOD}
Growing large high-quality single crystals is very important for reliable physical property measurements. Since all high-$T_c$ superconductors melt incongruently, i.e. decompose before melting, direct crystallization from the melt (\textit{melt growth}), such as occurs in the Bridgman method or the Czochralski method, can not be used because it results in some other phase. The alternative is \textit{solution growth}, in which the constituents of the material to be crystallized are dissolved in a suitable solvent and crystallization occurs as the solution becomes critically supersaturated \cite{elw}. Among \textit{solution growth} methods, the most popular two in the crystal growth of the high-$T_c$ superconductors are the flux method (slow cooling method) and the traveling solvent floating zone (TSFZ) method.

The principle of the flux method is shown in Fig. 2 as (a) a phase diagram and (b) a practical image. In this method, the supersaturation is obtained by cooling. Continuous crystal growth occurs because the solubility decreases with decreasing temperature as shown in Fig. 2(a). The flux method is simple and inexpensive and is thus widely used in preparing new phases. On the other hand, the grown crystals are often contaminated from the crucible used and it is difficult to grow large crystals. More seriously, crystals cannot practically be obtained in a system where the solubility difference available is very small (i.e., the liquidus line is very steep and/or narrow).

The principle of the TSFZ method is shown in Fig. 3 as (a) a phase diagram and (b) a practical image. A proper amount of solvent with the composition "s" is set between a feed rod and a seed crystal, melted, and then they are slowly moved downward [Fig. 3(b)]. Consequently, crystals with the composition of "T" are precipitated on the seed crystal through a solid-liquid reaction between "T" and "s". The solvent composition "s" is maintained unchanged because the solute exhausted by the crystallization is continuously supplied (transported) from the feed by melting in the steady growth state. The supersaturation, in this case, is obtained by nutrient (solute) transport. Therefore, crystal growth continues at one point on the temperature-composition phase diagram as shown in Fig. 3(a). Consequently, the TSFZ method allows us to grow large single crystals even if the liquidus line (crystallization field) is very narrow. For the same reason, this method is especially useful for the preparation of homogeneous solid solutions and homogeneously doped crystals \cite{wat1}. Furthermore, the crystal growth is performed in a free space, i.e., a crucible is not used, so the obtained crystals are free from impurities. It is difficult, however, to keep the molten zone stable, because it is supported only by surface tension. For example, the molten zone for various reasons, sometimes drops down, soaks into the feed rod, and bubbles by many reasons. Therefore, the applicability of the TSFZ method has generally been thought to be narrow. We choose the TSFZ method for the crystal growth of Bi$_{2}$Sr$_{2}$Ca$_{n-1}$Cu$_{n}$O$_{2n+4+\delta}$ ($n$ = 2, 3) systems, because, in principle, it makes growth of the large high-quality single crystal possible.   

\subsection{CONCEPT OF THE I-TSFZ METHOD}        

As mentioned above, some materials are difficult to obtain in single crystal form, mainly because they have only narrow liquidus lines. With narrow liquidus lines, the supersaturation is small, which makes it difficult to grow large crystals by standard crystal growth methods such as the flux method. Figure 4 shows a schematic phase diagram of the Bi family. Obviously, Bi$_{2}$Sr$_{2}$Ca$_{n-1}$Cu$_{n}$O$_{2n+4+\delta}$ ($n$ = 2, 3) systems, especially the $n$ = 3 compound, Bi-2223, has a narrow liquidus line. To overcome this serious problem of the insufficient supersaturation, we developed the I-TSFZ method with two features \cite{fuj1,wat2}.

One feature of the I-TSFZ method is the very slow growth rate adopted. The driving force for crystal growth is provided by supersaturation \cite{elw}. The maximum growth rate increases as the level of supersaturation increases. Therefore, if the supersaturation level is very low, we should use a very slow growth rate so as not to exceed the maximum growth rate. Furthermore, these Bi compounds (Bi-2212 and Bi-2223) are known to grow very anisotropically; i.e., the growth rate along the c-axis is extremely low compared with that along the $a$- or $b$-axis. Probably because of this anisotropic growth rate, single crystals of these compounds are usually thin ($\le$ 0.1 mm and $\le$ 0.003 mm for Bi-2212 and Bi-2223, respectively) along the $c$-axis. The very slow growth rate is expected to allow sufficient time for growth along the c-axis and thus increase the crystal thickness.

The other feature is the steep temperature gradient at the solid-liquid interface. In the TSFZ method, the whole temperature range covering the liquidus line is not necessarily available because the molten zone sometimes drops depending on its size, viscosity, and other factors. We found that a steep temperature gradient provides comparatively flattened solid-liquid interfaces for both feed and grown crystal sides, which makes it possible to reduce the length of the molten zone. It is considered that the molten zone becomes stable if its size is reduced because it is supported only by surface tension. Therefore, the steep temperature gradient is expected to stabilize the molten zone and thereby make it possible to control the growth temperature precisely. On the other hand, it is important in the TSFZ method that the solvent composition and quantity be kept exactly at the specific optimal values. Fortunately, in the TSFZ growth of Bi-based superconductors, the solvent composition can be self-adjusted during growth if the molten zone is sufficiently stable against composition and temperature fluctuation. The steep temperature gradient is expected to enhance the self-adjustability by stabilizing the molten zone. In addition, more importantly, the supersaturation level increases with increasing temperature gradient. As noted above, supersaturation is the driving force of crystal growth in solution growth. Therefore, the steep temperature gradient is expected to enhance the maximum growth rate, which was considered to be relatively lower in the system having narrow liquidus line (crystallization field). 

Noteworthy is the intimate relation between the conditions for stable crystal growth and the I-TSFZ method. The instability of a planar crystal surface growing in a doped melt or in a solution has often been discussed in terms of \textit{constitutional supercooling} \cite{til1}. As the crystal grows, solvent is rejected at the crystal surface and so the solute concentration immediately ahead of the interface becomes appreciably lower than that in the bulk of the solution [Fig. 5(a)]. This accumulation of solvent near the interface results in a depression of the equiliblium liquidus temperature $T_L$ as shown in Fig. 5(b). If the actual temperature distribution in the solution is as shown in line "A" of Fig. 5(b), any protuberance on the interface will tend to grow since it will experience higher supercooling. The region ahead of the interface will then become unstable, which results in the so-called cellular growth with multi-domained crystals and precipitations in between them. Stable growth can occur if the temperature gradient at the interface is increased as in "B" of Fig. 5(b), in which case the actual temperature everywhere ahead of the interface is higher than the liquidus temperature. The condition for \textit{constitutional supercooling} when it is applied to the \textit{solution growth} is expressed by Tiller \cite{til2} as
\begin{equation}
v \textgreater D_o\frac{dT}{dx}\bigg/\sum_{i=1}^{j}\biggl(\frac{m_i(k_i^*-1) n_i}{D_i/D_o}\biggr)
\end{equation}
where $v$ is the growth rate, $D_o$ the diffusion coefficient of the solvent, and $m_i$, $k_i^*$, $n_i$, and $D_i$ are the slope of the liquidus curve, the effective partition coefficient, the concentration in the solution, and the diffusion coefficient of the solute constituent $i$, respectively. The effective partition coefficient is defined by
\begin{equation}
k_i^* = \Bigl(\frac{n_c^i}{n_{sn}^i}\Bigr)_{x=0}
\end{equation}
where $n_c^i$ and $n_{sn}^i$ are the concentrations of each solute constituent $i$ in the crystal and solution, respectively. From eq. (1), the very slow growth rate and the steep temperature gradient adopted in the I-TSFZ method is expected to work cooperatively to prevent the occurrence of \textit{constitutional supercooling}, which is a cause of cellular growth.

\subsection{Bi$_{2}$Sr$_{2}$CaCu$_2$O$_{8+\delta}$:Bi-2212}        
Single crystal growth of the Bi-2212 phase by the TSFZ method was first reported by Takekawa et al. \cite{takek}. For the crystal growth, they used an infrared convergence-type floating zone furnace with a 1500-W halogen lamp as the radiation source. The growth rate was 1 mm/h. The grown boule was an aggregate of large single crystal grains with growth direction along the a-axis. Although the size was sufficient in $a$- and $b$-axis direction (several mm), that in the $c$-axis direction (thickness) was limited below 0.1 mm.

To increase the crystal thickness, we decreased the growth rate to 0.5 mm/h. In addition, we increased the temperature gradient by using 300-W halogen lamps (A furnace equipped with two halogen lamps was used) in place of the standard 1500-W lamps. The temperature gradient was about 370$^\circ$C/cm, which is two times larger than that with 1500-W lamps. Actually, the length of the molten zone is reduced from 4.0-4.5 mm for 1500-W halogen lamps to 3.5-4.0 mm. Compositions of Bi$_{2.1}$Sr$_{1.9}$CaCu$_2$O$_{8+\delta}$ and Bi$_{2.2}$Sr$_{1.6}$Ca$_{0.85}$Cu$_{2.2}$O$_{8+\delta}$ were used for the feed and the molten zone, respectively. The samples were prepared from powders of Bi$_2$O$_3$, SrCO$_3$, CaCO$_3$, and CuO (all of 99.999\% purity) mixed in the desired ratio, and then calcined at 780$^\circ$C (feed) and 760$^\circ$C (molten zone) for 12 h with repeated regrindings. The calcined powders were again reground, placed into a rubber tube and then hydrostatically pressed under 2000 kg/cm$^2$. The pressed rods for the feed and for the molten zone were sintered at 860$^\circ$C and 830$^\circ$C, respectively, for 50 h. The sintered feed rod was premelted at a rate of 25 mm/h to prepare a dense feed rod. The feed rod was suspended at the bottom of the upper shaft, and a seed crystal (or just a sintered rod) was held at the top of the lower shaft. A proper amount of sintered rod (pellet) for the molten zone was set under the bottom of the feed rod and was melted to form a proper solid-liquid reaction. The upper and the lower shafts were counterrotated at a rate of 10.5 and 10.0 rpm, respectively. Then, the lower shaft was moved downward at a rate (growth rate) of 0.5 mm/h, while the upper shaft was moved at a rate of 0.4 mm/h. The growth atmosphere was a mixed gas flow of O$_2$ (20\%) and Ar (80\%).

Figure 6(a) shows an image of the whole grown boule. Crystals tend to become large as growth proceeds. Figure 6(b) shows an optically polarized micrograph of a cross section normal to the growth direction of nearly the last part of the grown boule. As can be seen in this figure, large crystals with the thickness of about 1 mm were obtained. Figure 6(c) shows an $ab$-plane image of such crystals cleaved from the boule. The $ab$-plane of the crystal was fairly large: the length along $a$-axis was over 5 mm and that along $b$-axis was 4.5 mm, which is almost the thickness of the grown boule. The main reasons for this successful result are that enough time was allowed for growth along the $c$-axis by the slow growth rate and that the $c$-axis maximum growth rate was enhanced by steep temperature gradient. These crystals have been used for the physical property measurements (section 3). A further increase of the temperature gradient, unfortunately, caused unstable crystal growth. Prior to our result, a similar or better result of 1.5 mm thick Bi-2212 crystal had been reported by Gu et al. \cite{gu}, who used a growth rate of 0.2 mm/h and a halogen lamp power of 500 W.

In an attempt to further enlarge the crystals, the growth rate was reduced to 0.05 mm/h with the other conditions fixed \cite{fuj2}. Figure 7(a) shows a cross-sectional image of the grown boule obtained with an optically polarized microscope. About one-third of the crystal boule (left part) was a single domain with a c-axis thickness of 1.5 mm, while the remaining part was heavily multi-domain. The obtained Bi-2212 crystals unfortunately contained a large number of Bi-2223 impurities. That such thick crystals were grown indicates that the slow growth rate is indeed effective for enlarging crystal size. However, the existence of heavily multi-domain region, on the other hand, suggests that another factor competes with crystal enlargement. Reflecting the extreme anisotropy in the growth rate between the $b$-axis and $c$-axis, the cross section becomes elliptical in shape as shown in the figure. This ellipsoidal cross-section of a crystal boule likely causes the molten zone to drop down along the minor axis of the ellipsoid. The situation is illustrated in Fig. 7(b). To prevent the molten zone dropping down, the remaining part [right part in Fig. 7(a)] may have self-assembled into a heavily multi-domain structure. Indeed, the solid-liquid interface of the remaining part is rather flat compared to that of the single-domain part [Fig. 7(c)]. Figure 7(d) and (e) show enlarged images of the solid-liquid interface of the single-domain part and that of the multi-domain part, respectively. A smooth interface is seen in Fig. 7(d), while a rough one is seen in Fig. 7(e), evidencing the single-domain growth and multi-domain growth, respectively. It should be noted that this multi-domain growth is not the cellular growth caused by \textit{constitutional supercooling}, which was discussed in section 2.2. Instead, it may be due to molten zone instability caused by the anisotropic growth rate. In this context, an interesting attempt has been reported by Watauchi et al. \cite{watau}, in which the anisotropic growth rate was controlled using an infrared heating furnace equipped with four mirrors. Taking CuGeO$_3$ and Sr$_{14}$Cu$_{24}$O$_{41}$ as examples, they showed that the ellipsoidal cross-section of a cylindrical crystal grown usually under an isotropic lamp power changed to a circular one when the crystals were grown under an anisotropic lamp power. 

\subsection{Bi$_{2}$Sr$_{2}$Ca$_2$Cu$_3$O$_{10+\delta}$:Bi-2223}        
In initial attempts at growing Bi-2223 phase, the KCl flux method was used \cite{bal,gor}, but the obtained crystals were very thin, with sizes of 1 x 1 x 0.003 mm$^3$. Obviously, this is due to the narrowness of the liquidus line. To improve the size and quality of this system, we adopted the I-TSFZ method \cite{fuj1}. As noted in the section 2.1, the TSFZ method is suitable for such single-crystal growth because the growth can be performed continuously at one point on the temperature-composition phase diagram. 

Crystal growth of the Bi-2223 phase by the I-TSFZ method was carried out with a growth rate of 0.05 mm/h and temperature gradient of 370$^\circ$C/cm using 300-W halogen lamps. For a feed rod, the nominal composition of Bi$_{2.1}$Sr$_{1.9}$Ca$_2$Cu$_3$O$_{10+\delta}$ was used. A solvent was not used. Other growth conditions were the same as for Bi-2212 (section 2.3). Figure 8(a) shows a feed rod premelted at a rate of 25 mm/h. Making this kind of dense feed rod with uniform thickness is indispensable for the successful crystal growth. Photographs [Fig. 8(b)] show actual crystal growth near the molten zone. They were taken once every week. The crystal growth was very stable for a long period (over two months). The final rod is shown in Fig. 8(c). From the last 1-cm part of the grown rod, large high-quality single crystals were harvested. Figure 8(d) shows an optically polarized micrograph of a cross section normal to the growth direction of this part. Although the grown boule consisted of many largely misoriented plate-like crystals of Bi-2223 as in the case of Bi-2212, large plate-like single crystals with dimensions of up to 4 x 2 x 0.1 mm$^3$ [Fig. 8(e)] could be cleaved from the upper right side [Fig. 8(d)]. Figure 9 shows the temperature dependence of the DC susceptibility of a crystal also cleaved from the last 1-cm part of the boule. There is a sharp superconducting transition ($\le$ 3 K) with $T_c$ = 105 K. After the sample was annealed in flowing oxygen at 600$^\circ$C, $T_c$ increased to the maximum value of 110 K, suggesting that the as-grown sample is in the underdoped region. The small step-like change in the shielding signal, $\Delta$M, around $T$ = 83 K may have been caused by the intergrowth of the Bi-2212 phase. However, $\Delta$M is less than 2$\%$ of the full volume of the demagnetization measured at 5 K. This assures that the obtained crystal is a nearly pure-phase Bi-2223 with less than 2$\%$ Bi-2212 impurity. A detailed structure analysis of this sample based on a four-circle x-ray diffraction measurement has recently been reported \cite{wat5}. Single crystals in other parts (other than the last 1-cm part) were small (typically 1 x 0.5 x 0.02 mm$^3$). They consisted mainly of the Bi-2223 phase with around a 20$\%$ Bi-2212 impurity phase. Since we did not use a solvent, the composition for the molten zone might not have been sufficiently suitable at the first to middle stage of the crystal growth. However, in the TSFZ growth of Bi-based superconductors, the composition for the molten zone is self-adjusted during growth. In the last 1-cm part of the grown boule, the composition might have been well stabilized at the optimal value for Bi-2223 growth.

We have successfully grown, for the first time, large high-quality single crystals of Bi-2223 by using the I-TSFZ method. A very slow growth rate and a steep temperature gradient at the solid-liquid interface made it possible to stabilize the growth conditions in the narrow crystallization field of Bi-2223. Recently, our result has been reproduced by several groups \cite{lia,fen}. Liang et al. \cite{lia} have emphasized that the growth rate affects the phase formation of Bi-2223 crystals. They have shown that the as-grown crystals obtained at a slow growth rate of 0.04 mm/h consist of a \textgreater 90$\%$ Bi-2223 phase, whereas as-grown crystals obtained at rates of 0.10 and 0.20 mm/h contain predominantly Bi-2212, Ca$_2$CuO$_3$, and a small quantity of the Bi-2223 phase.

\section{SOME PHYSICAL PROPERTIES}
\subsection{Bi$_{2}$Sr$_{2}$CaCu$_2$O$_{8+\delta}$:Bi-2212}
\subsubsection{PRECISE OXYGEN CONTROL}
Bi-2212 is a suitable system for studying the electronic phase diagram noted in the introduction, because its doping level can be controlled over a wide range by varying the oxygen content. However, it had been somewhat difficult to control the doping levels, especially into the underdoped region, and retain good sample quality. To overcome this difficulty, we developed a technique to control the oxygen content ($\delta$) of Bi-2212 crystals precisely \cite{wat3}. 

Figure 10 shows an equilibrium phase diagram of Bi-2212 single crystals for various $\delta$, which was obtained by a thermogravimetric measurement \cite{wat3}. According to the phase diagram, the oxygen content ($\delta$) of the crystals was controlled to the required values: the crystals were annealed in a quartz tube furnace under the corresponding oxygen partial pressures at 600$^\circ$C for 10 h, slowly cooled ($\approx$0.4 $^\circ$C /min) while maintaining the equilibrium oxygen pressures as a function of temperature to 300-400$^\circ$C, and then rapidly cooled. This process ensured both uniform oxygen content and minimum disorder in the crystal. Figure 11 shows several DC susceptibilities of the annealed samples as well as that of the as-grown one. The annealed samples show narrower superconducting transition ($\le$3 K) than the as-grown one (transition width $\approx$5 K), confirming the uniform oxygen content. The superconducting transition temperature ($T_c$) increased, reached a maximum, then decreased with increasing oxygen content ($\delta$). The $T_c$'s are plotted as a function of $\delta$ in Fig. 12. Fig. 13 shows the experimental setup for the annealing. For the control of the oxygen partial pressures, a mixed gas flow of O$_2$ and Ar at a total amount of 5 l/min was used in the higher oxygen pressure region ($P_{O2}$ $\ge$ 10$^{-4}$ atm). In the lower oxygen pressure region ($P_{O2}$ \textless 10$^{-4}$ atm), the pressure around the sample was controlled by tuning the conductance of the evacuation path using a slotted valve, as well as by tuning the oxygen flow while continuously evacuating the quartz tube. A personal computer controlled the whole system, i.e. both the oxygen partial pressures and the temperatures. Heavily overdoped samples with $T_c$ $\approx$ 60 K ($\delta$ $\approx$ 0.3) were made separately by high O$_2$ pressure (400 atm) annealing at 500 $^\circ$C for 50 h using a hot isostatic pressing (HIP) furnace.

\subsubsection{$\rho_{a}$, $\rho_{c}$ and $\chi$}
Figure 14 shows the temperature dependence of in-plane resistivity $\rho_{a}$ for Bi-2212 single crystals with various oxygen contents ($\delta$) \cite{wat3}. The overall slope $d\rho_{a}/dT$ decreases with increasing $\delta$. In a simplest model, resistivity $\rho(T)$ is expressed as 
\begin{equation}
\rho(T) = \frac{m^*}{ne^2\tau(T)}
\end{equation}
where $m^*$ is the effective mass, $n$ the carrier concentration, $e$ the charge,  and $\tau$ the carrier lifetime. The monotonic decrease of $d\rho_{a}/dT$ assures that the carrier concentrations are indeed increased by the oxygenation process. A typical $T$-linear behavior and a slightly upward curvature of $\rho_{a}$ are also seen in an optimally doped ($\delta$=0.255, $T_c$=89 K) and an overdoped sample ($\delta$=0.28, $T_c$=81 K), respectively. For the underdoped samples ($\delta$$\textless$0.25), $\rho_{a}$ deviates from high-temperature $T$-linear behavior at a characteristic temperature $T^*_{{\rho}_a}$ (shown by the arrow in Fig. 14), and decreases rapidly with decreasing temperature. In the  YBa$_2$Cu$_3$O$_{6+\delta}$ system, this behavior has been interpreted as being due to a decrease in spin scattering, which is caused by the pseudogap opening \cite{ito}. $T^*_{{\rho}_a}$ in Bi-2212 is estimated as 220, 200, and 175 K for $\delta$ = 0.2135, 0.217, and 0.22, respectively.

Figure 15(a) shows the temperature dependence of out-of-plane resistivity  $\rho_{c}$ for a Bi-2212 single crystal (sample I) with various oxygen contents \cite{wat3}. A typical semiconductive temperature dependence is observed in a wide doping range. It should be noted that this semiconducting behavior of $\rho_{c}$ is quite anomalous in that the temperature dependence of $\rho_{a}$ and $\rho_{c}$ is totally different. In general, in a metal, the temperature dependence of $\rho_{a}$ and $\rho_{c}$ is fundamentally the same ($\rho_{a}(T)$/$\rho_{c}(T)$ = constant) even if the system is two-dimensional (2D), whereas their magnitude differs ($\rho_{a}(T)$/$\rho_{c}(T)$ $\neq$ 1), reflecting the anisotropic effective mass $m^*$. Although many models have been proposed to explain semiconductive $\rho_{c}$, there was no consensus \cite{coo}. Figure 15(b) shows $\rho_{c}$ for another Bi-2212 single crystal (sample II), focusing on its optimum and overdoped behavior \cite{wat4}. The $\rho_{c}$ in an optimally doped sample ($\delta$=0.25, $T_c$=89 K) shows semiconductive behavior in all temperature regions measured. With increasing $\delta$, the overall magnitude of $\rho_{c}$ decreases and the characteristic temperature $T^\ast_{\rho_c}$ [shown by the arrow in Fig. 15(b)] for the onset of upturn appears. $T^\ast_{\rho_c}$ is estimated as 185, 165, and 130 K for $\delta$=0.26, 0.27, and 0.28, respectively. For the heavily overdoped sample ($\delta$$\approx$0.3, $T_c$=60 K), $\rho_{c}$ shows no upturn behavior, suggesting the pseudogap completely vanishes. 

Figures 16(a) and (b) show the temperature dependence of magnetic susceptibilities ${\chi}_{c}$ $(H{\parallel}c)$ and ${\chi}_{ab}$ $(H{\perp}c)$, respectively, of the B-2212 single crystals with various oxygen contents \cite{wat4}. The overall magnitude of ${\chi}_{c}$ and ${\chi}_{ab}$ increases with increasing $\delta$. It is presumed that the magnetic susceptibilities can be expressed as
\begin{equation}
\chi_\alpha(T) = \chi_{core} + \chi_\alpha^{VV} + \sum_{i=1}^{n}g(i)_\alpha^2\chi_{spin}(i,T)
\end{equation}
where $\alpha$ denotes $c$ or $ab$, $\chi_{core}$ denotes the sum of the core diamagnetism for the ions present, $\chi_\alpha^{VV}$ is the anisotropic Van Vleck paramagnetism (orbital contribution from the Cu$^{2+}$ ion), $\chi_{spin}(i,T)$ the i-th component of the spin susceptibility for the doped CuO$_2$ plane (For generality, we assumed $n$-components for the spin susceptibility), and $g(i)_\alpha$ denotes their anisotropic $g$-factors. The temperature and doping dependence of ${\chi}$ mainly come from the spin susceptibility. The susceptibilities in the underdoped ($\delta$=0.22, $T_c$=82 K) and optimally doped ($\delta$=0.25, $T_c$=89 K) samples monotonically decreases with decreasing temperature. For the slightly overdoped sample ($\delta$=0.26, $T_c$=86 K), they are almost constant at high temperatures and gradually decrease from a characteristic temperature $T^\ast_{\chi}$ [shown by the arrow in Fig. 16(a)] upon cooling. A further increase in $\delta$ causes a shift of $T^\ast_{\chi}$ to lower temperatures, whereas, above $T^\ast_{\chi}$, the susceptibilities prominently decrease with increasing temperature. $T^\ast_{\chi}$ is estimated as 200, 175, and 135 K for $\delta$=0.26, 0.27, and 0.28, respectively. The difference in the absolute value between $\chi_c$ and $\chi_{ab}$ may be caused by the anisotropic Van Vleck paramagnetism, which is almost completely $T$-independent \cite{joh2}.

\subsubsection{ESTIMATION OF DOPING LEVEL FROM $\rho_{c}(T)$}
The above systematic measurement for $\rho_{c}(T)$ presented us with a new method for estimating the doping levels $p$ of Bi-2212. Bi-2212 is now a key material for studying the physics of high-$T_c$ superconductivity, since several important spectroscopic measurements, such as angle resolved photoemission spectroscopy (ARPES) \cite{din,mar}, scanning tunneling spectroscopy (STS) \cite{ren,mat1} and interlayer tunneling spectroscopy (ITS) \cite{suz1,ana}, can be performed on this material owing to its clean surface and highly electronic 2D nature. Determining the doping levels ($p$) of a sample is, therefore, an important issue. In general, $p$ is estimated from $T_c$ using the empirical relation \cite{tal}
\begin{equation}
\frac{T_c}{T_{c,max}} = 1-82.6(p-0.16)^2
\end{equation}
where $T_{c,max}$ is the maximum $T_c$ value of the Bi-2212 sample. However, in this case, controlling the $p$ in a wide range including $T_{c,max}$ is a prerequisite, but is not always possible. Alternatively, we found that $\rho_{c}(T)$ provides a simple and reliable estimation for $p$.

Figure 17 plots, for several samples (I, II, and III), ratios of peak value, $\rho_{c,peak}$, of $\rho_{c}$ just above $T_c$ to $\rho_{c}$ at room temperature, $\rho_{c}(RT)$, together with their $T_c$, as a function of doping levels ($p$). Here, excess oxygen $\delta$ was converted to $p$ using the above empirical relation. Although the absolute value of $\rho_{c}(T)$ scatters from sample to sample, the ratio $r = \rho_{c,peak}/\rho_{c}(RT)$ scales in one curve
\begin{equation}
p = 0.024 + \frac{0.993}{r + 4.696}.
\end{equation}
This scaling function enables us to estimate $p$ from a simple $\rho_{c}(T)$ measurement. The advantages of this method are that it is possible to estimate $p$ using one sample and the estimation is independent of the sample quality. This method was actually used in an ITS measurement \cite{suz1}. 
 
\subsubsection{PSEUDOGAP PHASE BOUNDARY}
As noted in the introduction, there is still no consensus on when (at what temperatures) the pseudogap opens. This is mainly because of our incomplete understanding of the $T$ dependence of the uniform susceptibility $\chi$. For example, in the La$_{2-x}$Sr$_x$CuO$_4$ system, it is known that $\chi$ first increases, takes a broad maximun, and then decreases from some characteristic temperature ($T^\ast_{\chi}$) with decreasing temperature and $T^\ast_{\chi}$ decreases with increasing doping (Sr content) \cite{joh1}. Our observation for Bi-2212 (Fig. 16) is qualitatively similar. D. C. Johnston interpreted the $\chi$ of the La$_{2-x}$Sr$_x$CuO$_4$ system as the sum of two-dimensional (2D) $S$=1/2 square-lattice Heisenberg antiferromagnetism and $T$-independent Pauli paramagnetism \cite{joh1}. Later, there was proposed a model based on the mode-mode coupling theory of spin fluctuations in 2D metals with a technically nested Fermi surface, which explains the $T$ dependence as a crossover from the localized to itinerant spin fluctuations \cite{miya}. In those pictures, the gradual decrease in $\chi$ is not attributed to the formation of the pseudogap. On the other hand, J. W. Loram et al. have shown from susceptibility and high-resolution specific heat measurements that the pseudogap develops below $T^\ast_{\chi}$ in the YBa$_2$Cu$_3$O$_{6+\delta}$ \cite{lor1} and La$_{2-x}$Sr$_x$CuO$_4$ systems \cite{lor2}. Similar results have been reported from NMR Knight shift measurements \cite{will1}. Up to now, we have not had a unified description of the $T$/doping dependence of $\chi$, that is, that the behavior is pseudogap-like at low dopings and Curie-like at high dopings.

Figure 18 plots the above obtained $T^\ast_{\rho_a}$, $T^\ast_{\rho_c}$ and $T^\ast_{\chi}$ together with several other characteristic temperatures, including $T_c$, as a function of carrier concentration $p$ \cite{wat4}. Here again $p$ was estimated from $T_c$ using the empirical equation. $T^\ast_{\chi}$ and $T^\ast_{\rho_c}$ coincide for each $p$. This strongly suggests that they have the same origin. Unlike YBa$_2$Cu$_3$O$_{6+\delta}$, which may change from 2D to 3D upon doping, Bi-2212 stays in 2D for all doping levels. We have previously shown that the $\rho_c$ in the Bi-2212 system is governed by a simple tunneling process \cite{wat1}. Then, $\rho_c$ should reflect the in-plane density of states (DOS) through the equation \cite{naga1}
\begin{equation}
\rho_{c} = \frac{A}{t_c^2{\int}_{-\infty}^{\infty}N^2(\varepsilon)(-{\partial}f(\varepsilon)/\partial\varepsilon)d\varepsilon}
\end{equation}
where A is a constant, $t_c$ the hopping probability in the c-axis direction, $N(\varepsilon)$ the DOS, and $f(\varepsilon)$ the Fermi function. For the spin susceptibility, we assume simply the Pauli paramagnetism
\begin{equation}
\chi_{pauli}=\mu_B^2{\int}_{-\infty}^{\infty}N(\varepsilon)(\frac{-{\partial}f(\varepsilon)}{\partial\varepsilon})d\varepsilon
\end{equation}
where $\mu_B$ is the Bohr magneton. Since the magnitude of Pauli paramagnetism also reflects the DOS, the coincidence of $T^\ast_{\chi}$ and $T^\ast_{\rho_c}$ simply indicates that the pseudogap opens at these temperatures (line "B" in Fig. 1). We have confirmed this pseudogap opening at $T^\ast_{\chi}$/$T^\ast_{\rho_c}$ by scanning tunneling spectroscopy (STS) \cite{mat2} (Some of the data is shown as $T^\ast_{tunnel}$ in Fig. 18). This result contradicts the explanation where the gradual decrease in $\chi$ is attributed to the development of AF-spin correlations, and the long-debated origin of the semiconductive $\rho_c(T)$ is now clear: it is the pseudogap opening. 

Our results here may conflict with NMR measurement \cite{ish} and the ARPES \cite{din}, which show that the pseudogap opens at around $T^\ast_{\rho_a}$ (line "A" in Fig. 1). Their characteristic temperatures, ($T^\ast_K$ and $T^\ast_{ARPES}$), are also shown in Fig. 18. We could not observe any discontinuous change in temperature dependence or anisotropy of the susceptibility at around line "A". We point out, however, that if $T^\ast$ in \cite{ish} is defined as the temperature at which 1/$T_1T$ departs from high-temperature Curie-Weiss behavior, it will come close to our pseudogap phase boundary "B". The leading edge analysis of ARPES may be too simple to detect the pseudogap opening temperature precisely, since the spectra broadens a lot due to both a high temperature measurement and a large pseudogap energy. 

The $\rho_c$ and $\chi$ may be particularly sensitive to the onset of the pseudogap opening. A saddle point is observed for many high-$T_c$ superconductors in the band dispersion at ($\pi$,0) \cite{shen}. The band near ($\pi$,0) is so flat that it creates a large peak [van Hove singularity (vHs)] in the DOS. Therefore, the DOS is concentrated near the ($\pi$,0) region. In addition, a recent ARPES study has shown that the pseudogap starts to open from the ($\pi$,0) direction \cite{nor1}. These situations explain why $\rho_c$ and $\chi$ are sensitive to the pseudogap opening. Further, a recent band calculation has shown that the hopping probability $t_c$ in the $c$-axis direction depends on momentum as $t_c \propto (cosk_xa - cosk_ya)^2$: $t_c$ takes maximum at ($\pi$,0). This momentum-dependent $t_c$ will enhance the sensitivity of $\rho_c$ to the pseudogap opening. On the other hand, $\rho_a$ may be less sensitive to the pseudogap opening.  The carriers around ($\pi$,0) are exposed to strong scattering ("hot spot"), while those around ($\pi/2$,$\pi/2$) are not scattered very much ("cold spot"). Consequently, carriers at the "cold spot" mainly contribute to $\rho_a$. When the pseudogap opens from the "hot spot", $\rho_a$ may not practically be affected. $\rho_a$ begins to feel the pseudogap opening when it reaches the "cold spot". However, the decrease in the $\rho_a$ means that the pseudogap is not the type that simply reduces Fermi surface area, such as charge-density-wave formation, but the type that only reduces the scattering rate, such as opening a gap in the spin excitation spectrum.

The experimentally obtained pseudogap phase boundary ($T^\ast_{\chi}$, $T^\ast_{\rho_c}$ and $T^\ast_{tunnel}$ in Fig. 18, "B" in Fig. 1) is not smoothly connected to the $T_c$ boundary of the heavily overdoped state; rather, it seems to cross the $T_c$ boundary at a slightly overdoped level and approach the quantum critical point at around $p$=0.20 ($T$=0) \cite{var,sach}. This suggests that the pseudogap is not simply a precursor of the superconducting gap \cite{ren}, but it is a manifestation of some quasi-order that competes with superconductivity \cite{var,sach}. Recently, several spectroscopic studies heve provided evidence supporting this view \cite{mat1,suz1,mat2,kras,mat3,alff,suz2,mat4}. Tunneling spectroscopy by using grain boundary junctions \cite{alff} showed, for the electron-doped Pr$_{2-x}$Ce$_x$CuO$_{4-y}$ and La$_{2-x}$Ce$_x$CuO$_{4-y}$, that a pseudogap state appears for the ground state when their superconducting state is destroyed by applying a magnetic field. This is important in that they established the pseudogap phase boundary $T^*$ within the superconducting region. Similar results have been reported for hole-doped Bi-2212 by using ITS \cite{kras,suz2}. In the YBa$_2$Cu$_3$O$_{6+\delta}$ case, $T^\ast_{\rho_c}$ \cite{tak} and $T^\ast_{\chi}$ \cite{lor1} indicate that the pseudogap vanishes at the optimum doping level, suggesting the boundary is near "A". However, even in this case, the true boundary sits at somewhat higher temperatures than line "A" of Fig. 1 and eventually crosses the $T_c$ boundary at the optimum doping. In this sense, YBa$_2$Cu$_3$O$_{6+\delta}$ is not exceptional. We speculate that the cause for somewhat lower pseudogap phase boundary of YBa$_2$Cu$_3$O$_{6+\delta}$ is the 3D nature specific to this system \cite{sem}.

\subsubsection{TEMPERATURE/DOPING DEPENDENCE OF $\chi$}
In order to analyze the $T$ dependence of anisotropic magnetic susceptibilities quantitatively, we plotted ${\chi}_{c}(T)$ vs. ${\chi}_{ab}(T)$ with an implicit parameter $T$ for several doping levels (Fig. 19) \cite{wat4}. In the analysis, we assume that the $T$ dependence comes from  spin susceptibilities of the doped CuO$_2$ plane [eq. (4)]. Each of the ${\chi}_{c}(T)$ vs. ${\chi}_{ab}(T)$ plots [Figs. 19(a), (b), and (c)] shows the same linear relation, i. e., $\chi_{c}\propto1.6\chi_{ab}$, except near $T_c$ ($\le$140 K). This implies that the spin susceptibility has a single component with the ratio of anisotropic $g$-factors, $(g_c/g_a)^2$=1.6. The Curie-like explanation for the ${\chi}$ of overdoped samples may therefore be excluded because it is unlikely that the localized paramagnetic centers, which could have appeared additively in the overdoped state, would accidentally have the same anisotropy as the main spin component. This single-component hypothesis is consistent with NMR experiments \cite{all1,all2} and ensures that we are actually measuring the intrinsic spin susceptibilities. The anisotropy ratio coincides with that for La$_{2-x}$Sr$_x$CuO$_4$ \cite{ter}, but is somewhat larger than Johnston's result for Bi-2212 \cite{joh2} and the result for YBa$_2$Cu$_3$O$_{6+\delta}$ \cite{all1,all2}. We believe our anisotropy ratio in the Bi-2212 is more accurate than in \cite{joh2}, since our value was directly obtained by the ${\chi}_{c}(T)$ vs. ${\chi}_{ab}(T)$ plot whereas, in \cite{joh2}, it was estimated assuming a simple model. The nonexistence of the CuO chain structure may explain the difference between Bi-2212 and YBa$_2$Cu$_3$O$_{6+\delta}$. In Figs. 18 and 19, we included the temperatures at which the deviation from the linear scaling becomes evident, $T_{scf}$. Since the diamagnetic superconductive fluctuation effect mainly appears on ${\chi}_{c}(T)$ in the 2D system, $T_{scf}$ is considered to be the onset temperature of superconductive fluctuation. Indeed, the diamagnetic component of the superconductive fluctuation $\delta\chi_{dia}$ follows the power law,  $\delta\chi_{dia} \propto \epsilon^{-2.3}$, where $\epsilon = ln(T/T_c)$, and gives the same magnitude independent of the doping level \cite{mat2}. This observation justifies our interpretation of $T_{scf}$. It should be noted that the superconductive fluctuation is independent of the pseudogap behavior and actually additive in the pseudogap region (Figs. 18 and 19).

As already mentioned, we consider that the $T$ dependence of $\chi$ for underdoped samples is caused by the pseudogap opening, which has been evidenced by STS \cite{mat1,mat2,mat3,ren,miyak}. Then, it is natural to consider that the Curie-like behavior of $\chi$ for overdoped samples is also caused by a DOS effect, since the system is expected to be more Fermi-liquid-like upon doping. One such possibility is van Hove singularity (vHs), whose existence has been shown by ARPES at somewhat below the Fermi energy ($\varepsilon_F$) for a slightly overdoped sample \cite{shen}. The vHs grows and approaches the $\varepsilon_F$ with increasing doping. Such doping-dependent vHs has actually been observed by our STS \cite{mat2}. Figure 20 shows the tunneling DOS taken at slightly above $T_c$ for various $\delta$'s. In the heavily overdoped sample with $\delta$ = 0.3, the DOS shows a clear peak (vHs) at the Fermi level. Then the peak shifts toward lower energy and rapidly broadens with decreasing doping level. For the sample with $\delta \le 0.27$, the pseudogap develops as a dip structure at the Fermi level. The dip structure becomes more pronounced when the doping level is decreased. Here, we assumed simply the Pauli paramagnetism $\chi_{pauli}$ for the single-component spin susceptibility [eq. (8)]. The  $\chi_{pauli}$ reflects the thermally averaged DOS near $\varepsilon_F$. Therefore, the anomalous energy-dependent DOS (pseudogap and vHs) and its doping evolution (Fig. 20) with the assumption of $\chi_{pauli}$ will explain the temperature/doping dependence of $\chi$ observed. This kind of interpretation was first proposed by Loram et al. \cite{lor2} and later asserted by Williams et al. \cite{will2}. Our data here support their interpretation from a different point of view.

The assumption of the Pauli paramagnetism for the spin susceptibility enables us to estimate the pseudogap value \cite{wat4}. Here in the underdoped state, the energy independent background DOS, $N_0$, was assumed (The vHs sits far below $\varepsilon_F$ and is smeared out). For the pseudogap, the BCS-type DOS (Dynes formula) with the d-wave symmetry gap was assumed: 
\begin{equation}
\frac{N(\varepsilon)}{N_0}=\int_0^{2\pi}\biggl(\frac{d \theta}{2\pi}\biggr)Re\biggl (\frac{|\varepsilon-i{\Gamma}|}{\sqrt{(\varepsilon-i\Gamma)^2-(\Delta\cos(2\theta))^2}}\biggr)
\end{equation}
where $\Gamma$ is the carrier scattering rate, assumed here to be 2$k_B$T in a consistent way with the transport data, and $\Delta$ is the maximum d-wave gap, the temperature dependence of which was neglected. The isotropic component of the spin susceptibilities, $\chi^{iso}_{spin}$, was obtained by averaging the anisotropic spin susceptibilities, i.e., $\chi^{iso}_{spin}$=(2/3)$\chi^{ab}_{spin}$+(1/3)$\chi^{c}_{spin}$, where $\chi^{ab}_{spin}$ and $\chi^{c}_{spin}$ were obtained using the values in the literature for $\chi_{ab}^{VV}$ and $\chi_{c}^{VV}$ \cite{joh2}. Then we can fit the experimental spin susceptibility $\chi^{iso}_{spin}(T)$ by the calculated ones with a free parameter $\Delta$ as shown in Fig. 21. The pseudogap value $\Delta$ was estimated as 65$\pm$3 and 48$\pm$3 meV for the underdoped ($\delta$=0.22) and optimum doped ($\delta$=0.25) samples, respectively. These pseudogap values coincide with our direct STS measurements \cite{mat2} and, possibly, with the "high energy gap" suggested by photoemission spectroscopy \cite{ron,ino}, but are larger than the superconducting gap (20 meV$\le$$\Delta$$\le$45 meV) obtained by ARPES/STS \cite{din,ren,mat2,miyak}. These different gap values between the pseudogap and the superconducting gap suggest again that the two gaps are different in origin.

\subsection{Bi$_{2}$Sr$_{2}$Ca$_2$Cu$_3$O$_{10+\delta}$:Bi-2223}
\subsubsection{DOPING DEPENDENCE OF $T_c$}
We controlled the oxygen content $\delta$ by annealing a sample while varying Ar and O$_2$ gas flow ratio and/or temperatures. The annealing conditions for Bi-2223 samples used are ($a$) O$_2$ 5x10$^{-3}$ torr, 600$^\circ$C; ($b$) O$_2$ 0.01$\%$, 600$^\circ$C; ($c$) O$_2$ 0.1$\%$, 600$^\circ$C; ($d$) O$_2$ 1$\%$, 600$^\circ$C; ($e$) O$_2$ 10$\%$, 600$^\circ$C; ($f$) O$_2$ 600$^\circ$C; ($g$) O$_2$ 500$^\circ$C; ($h$) O$_2$ 400$^\circ$C; and ($i$) HIP O$_2$ 400 atm, 500$^\circ$C (These descriptions are used in all figures). The superconducting transition temperature $T_c$ was defined as the onset of the Meissner effect.

Figure 22(a) shows the normalized transition temperature $T_c$ plotted against the relative change of the $c$-axis length from that of the optimum doped sample "$f$" \cite{fuj3}. The relation between the $T_c$ and $c$-axis length of Bi-2212 is also plotted for comparison. The $T_c$ and c-axis length are 89 K and 30.864 $\AA$ for optimally doped Bi-2212, and 108 K and 37.119 $\AA$ for Bi-2223 (sample $f$), respectively. The $c$-axis length monotonically decreases with increasing $\delta$ both in Bi-2212 and Bi-2223, indicating that oxygen is actually incorporated into crystals. In the case of Bi-2212, reflecting the bell-shaped doping dependence of $T_c$, which is common behavior in monolayer or bilayer cuprates, $T_c$ increases with decreasing $c$-axis length, reaches its maximum, and then decreases with decreasing $c$-axis length. In the case of Bi-2223, we see a similar increase of $T_c$ with decreasing $c$-axis length. However, $T_c$ keeps its maximum value as the $c$-axis length further decreases. This kind of anomalous doping dependent $T_c$ was first reported in the four-layered compound (Cu$_{1-x-\delta}C_x$)Ba$_2$Ca$_3$Cu$_4$O$_y$ (x=0-0.3) \cite{ogi}.

To confirm the carrier doping in the constant-$T_c$ region, we measured the doping dependence of the thermopowers $S$ [Fig. 22(b)] \cite{fuj3}. The magnitude of the thermopower monotonically decreases with increasing $\delta$. This result clearly shows that the carrier was doped continuously even in the constant-$T_c$ region. The room-temperature thermopower is considered to be a universal measure of the doping level \cite{obe}. By using this measure, the (average) doping level of sample $f$ can be considered the "optimal" doping. Then, the sample $i$ would be slightly overdoped with carrier concentration of about $p$ = 0.185. Thus, it would show $T_c$ of 102 K, if the carriers were homogeneously doped. Hereafter, we will refer to the constant-$T_c$ region as the overdoped region.

\subsubsection{$\rho_{ab}$ and $\rho_{c}$} 
The temperature dependence of in-plane resistivity $\rho_{ab}(T)$ with various $\delta$'s is shown in Fig. 23 \cite{fuj3}. The overall doping dependent behavior, such as the absolute values of $\rho_{ab}$ and the slopes $d\rho_{ab}/dT$, is similar to Bi-2212 \cite{wat3} except for $T_c$.  $T_c$ determined by zero resistivity agrees with those determined by magnetic susceptibility (inset of Fig. 23). A difference appears in the residual resistivity. At all doping levels, the samples show negative residual resistivity as indicated by the solid lines in Fig. 23. High-$T_c$ materials with $T_c$ larger than 100 K tend to show negative residual resistivity, although we do not know its relevance to $T_c$. The Bi-2223 also seems to belong to this class. The underdoped samples (denoted by $b$, $d$, $e$ and $f$) show a downward deviation from high-temperature $T$-linear behavior below a certain temperature $T^\ast_{\rho_ab}$, similarly to that of Bi-2212 \cite{wat3}. $T^\ast_{\rho_ab}$ increases with decreasing doping as 168, 192, 203, and 213 K for the samples $f$, $e$, $d$, and $b$, respectively, as indicated by arrows in Fig. 23.

Figure 24 shows the c-axis resistivity $\rho_{c}(T)$ for various $\delta$'s \cite{fuj3}. Similar to $\rho_{ab}$, we can see that $T_c$ is pinned at the maximum value in the overdoped region. The overall magnitude of $\rho_{c}$ decreases with increasing $\delta$, which is similar to Bi-2212 \cite{wat3,wat4}. Here, we can assume the tunneling model [eq. (7)] for $\rho_{c}(T)$ of Bi-2223 also. Then, the decrease in the absolute value of $\rho_{c}$ with $\delta$ would imply an increase in the in-plane DOS and semiconductive behavior would be attributed to the decrease of DOS caused by the pseudogap formation. The underdoped samples from $a$ to $f$ show semiconductive $\rho_{c}$ in all temperature regions measured, indicating that the pseudogap opens above room temperature. A difference appears in the overdoped region. We can see that the $T^\ast_{\rho_c}$ remains unchanged ($\approx$ 220 K) for doping levels higher than $g$, while the absolute value of $\rho_{c}$ continues to decrease.

\subsubsection{INEQUIVALENT HOLE DOPING AND CHARGE CONFINEMENT}
Figure 25 shows a schematic of the basic structure of the Bi-2223 phase \cite{wat5}. As can be seen in this figure, the Bi-2223 system has crystallographically inequivalent CuO$_2$ planes: outer CuO$_2$ planes with pyramidal (five) oxygen coordination and an inner CuO$_2$ plane with square (four) oxygen coordination. Then, there arises the possibility of inhomogeneous doping among inequivalent CuO$_2$ planes. On the other hand, in high-$T_c$ superconductors, there is a consensus that the sets of CuO$_2$ planes separated by the blocking layer are only weakly coupled and the interaction between them can be understood as a tunneling process. This is known as the charge confinement effect from the theoretical point of view \cite{and}. However, it is still unknown whether the confinement works for the CuO$_2$ planes within the unit cell. 

As shown in the previous section, we have observed in Bi-2223 that $T_c$ and characteristic temperature $T^*_{\rho_c}$ do not change in the overdoped region, while the magnitude of $\rho_c$ monotonically decreases and shows about half the value of an optimally doped sample ($f$) in the oxygen-rich state ($i$). The anomalous out-of-plane transport behavior indicates that there are more holes in the outer planes than in the inner plane. That is, the behavior can be understood by assuming that the holes are mostly doped in the outer CuO$_2$ planes and that the inner CuO$_2$ plane remains at the optimum doping level when the doping is increased. The decrease in $\rho_c$ can be attributed to the increased electrical coupling between the outer planes. It would be difficult to explain such a large decrease in $\rho_c$ if there were more holes in the inner plane than in the outer planes, since the $\rho_c$ would probably be dominated by the tunnel conductance between the outer planes separated by the blocking Bi$_2$O$_2$ layer. Bulk $T_c$ is determined by the always optimally doped inner plane, which has higher $T_c$ than the outer planes. In this case, bulk superconductivity is driven by a proximity effect. This inequivalent hole doping agrees with that observed in a four-layered compound (Cu$_{0.6}$C$_{0.4}$)Ba$_2$Ca$_3$Cu$_4$O$_{12+y}$ ($T_c$ = 117 K) \cite{Ctoku}. Our result shows that this inequivalent hole doping nature is generic in multilayered high-$T_c$ cuprates that have three or more CuO$_2$ planes in the unit cell.

On the other hand, $T^*_{\rho_c}$ pinning in the overdoped region, as well as semiconductive $\rho_c$, is also determined by the always optimally doped inner plane. Since the outer planes are overdoped, they do not solely show the pseudogap behavior. Then the observed pseudogap effect may come from the tunneling through inner planes. This strongly suggests that carriers in the normal state are confined to individual CuO$_2$ planes.

\subsubsection{WHY $T_c$ IS HIGHER IN TRIPLE-LAYERED COMPOUNDS} 
It is well established in mono- or bilayer high-$T_c$ superconductors that $T_c$ first increases, takes a maximum, and then decreases with increasing doping $p$ (Fig. 1). Although this dome-shaped relation between $T_c$ and $p$ is not completely understood, the model proposed by Emery and Kivelson \cite{eme} may be promising. In this model, the trade-off between two basic ingredients of superconductivity, i.e. pairing potential $\Delta$ and phase stiffness $\rho_s$, are considered. In conventional superconductors, $\Delta$ is much smaller than $\rho_s$, and thus $\Delta$ always limits $T_c$. However, in high-$T_c$ cuprates, the two energy scales are comparable. The $\Delta$ is maximal in the underdoped region and falls rapidly with increasing doping, whereas $\rho_s$ increases with increasing doping. Optimal $T_c$ is obtained at a crossover from a phase-ordering-dominated region (underdoped) to a pairing dominated region (overdoped). Extending this idea, Kivelson recently proposed a way to raise $T_c$ of high-$T_c$ superconductors \cite{kiv}. His idea is that higher $T_c$'s can be obtained in an array of coupled planes with different doped hole concentrations, such that a high pairing energy is derived from the optimally or underdoped plane and a large phase stiffness from the overdoped ones. We observed that, in the overdoped region of Bi-2223, outer planes are overdoped while an inner plane remains at near optimum doped (inequivalent hole doping). Here, the coupled system that Kivelsom assumed seems to be realized. We consider that Kivelson's scenario could possibly explain the higher $T_c$ of Bi-2223 \cite{fuj3}.

Recently, many spectroscopic measurements have been performed on the Bi-2223 single crystals \cite{fen,yama,sato1,matsui1,matsui2,sato2}. Fen et al. \cite{fen} observed by ARPES that the superconducting gap magnitude $\Delta$ and the relative coherence-peak intensity, which is regarded as a measure of the superfluid density, both scale linearly with $T_c$ for the optimally doped Bi family (Bi-2201, Bi-2212 and Bi-2223). Based on these observations, they argued that the higher $T_c$ of Bi-2223 is caused by the enhancement of both the pairing strength and phase stiffness, which is consistent with the original idea of Emery and Kivelson \cite{eme}. By using short-pulse interlayer tunneling spectroscopy (ITS), Yamada et al. \cite{yama} observed, for Bi-2223, a similar or a slight larger superconducting gap than that of Bi-2212, although they are skeptical about the scaling between $\Delta$ and $T_c$. They also reported doping dependence of $\Delta$, where $\Delta$ monotonically decreases upon doping even in the overdoped ($T_c$ constant) region, and they did not observe two gap structure in those region. This behavior is resonably explained in terms of McMillan's tunneling model for the proximity effect \cite{mcm}, provided one assumes a somewhat strong coupling between inner and outer planes. On the other hand, Sato et al. \cite{sato1} emphasized the scaling behavior. They showed that the characteristic energies not only of superconducting but also of pseudogap behavior are scaled with $T_c$ among the Bi family. From these observations, they implied that the effective superexchange interaction $J_{eff}$ of Bi-2223 is larger than that of the others (Bi-2212 or Bi-2201), consistently with spin mechanism for superconductivity. The research on the physics of triple-layered or more generally of multi-layered (the number of CuO$_2$ plane $n$ $\ge$ 3) high-$T_c$ superconductors, has just been launched. At present stage, the central issue of why $T_c$ is higher in those systems is an open question. 

\section{SUMMARY}
Using the I-TSFZ method, we have succeeded in growing large high-quality Bi-2212 and Bi-2223 single crystals. This method makes the best use of the advantage of the TSFZ method that crystal growth can be performed at one point on the temperature-composition phase diagram. The I-TSFZ method will be especially useful for materials that are difficult to grow. By the systematic transport and susceptibility measurements for those Bi-2212 and Bi-2223 single crystals, we found that (i) the pseudogap phase boundary crosses the $T_c$ boundary near optimal doping and approaches the quantum critical point (Fig. 26), suggesting that the pseudogap is not simply a precursor pairing, but a manifestation of some (hidden) quasi-order that competes or at least coexists with superconductivity; (ii) the semiconductive $\rho_c(T)$ is caused by the pseudogap opening; (iii) temperature/doping dependence of magnetic uniform susceptibility $\chi$ can be interpreted by the anomalous energy-dependent DOS [pseudogap and van Hove singularity (vHs)]; and (iv) charge confinement occurs at each individual CuO$_2$ plane. Furthermore, systematic transport measurements for Bi-2223 revealed that $T_c$ and $T^*_{\rho_c}$ become pinned when doping is increased in the overdoped region. This anomalous behavior is interpreted in terms of the inequivalent hole doping (an always optimal inner plane with overdoped outer planes) and the charge confinement. The combination of a largr superconducting gap $\Delta$ in the inner plane and a large superfluid density of the outer planes would explain the higher $T_c$ of Bi-2223 (Kivelson's model). Our finding suggests that both high $T_c$ and large phase stiffness, which measures the ability to carry a supercurrent, are compatible in multilayered high-$T_c$ cuprates. From this point of view, the multilayered system would be very promising for practical use. 

\begin{verbatim}
Acknowledgement 
We thank Prof. Minoru Suzuki for helpful discussions on the 
physics of high-Tc superconductivity. We thank Dr. C. Sekar 
for helpful discussions on crystal growth.

\end{verbatim}

\clearpage

Figure captions
\par
\noindent
\\
Fig. 1.
Schematic phase diagram of high-$T_c$ superconductors (for introduction).
\par
\noindent
\\
Fig. 2.
Principle of the flux method. (a) Phase diagram and (b) practical image. 
\par
\noindent
\\
Fig. 3.
Principle of the TSFZ method. (a) Phase diagram and (b) practical image. 
\par
\noindent
\\
Fig. 4.
Schematic phase diagram of Bi family (Bi-2201, Bi-2212 and Bi-2223).
\par
\noindent
\\
Fig. 5.
(a) Solute concentration $n_{sn}$ ahead of crystal growing in solution. (b) Equilibrium liquidus temperature $T_L$ ahead of crystal growing in solution. Temperature gradient smaller than "B" (for example "A") causes constitutional supercooling.
\par
\noindent
\\
Fig. 6.
(a) Obtained Bi-2212 boule. (b) Optical polarization micrograph of a transverse section of the grown boule. Large single crystal grain with thickness larger than 1 mm is seen in the central region (from bottom left to upper right region) of the micrograph. (c) Grown crystal cleaved from the boule.
\par
\noindent
\\
Fig. 7.
Crystal growth of Bi-2212 with a growth rate of 0.05 mm/h. (a) Optical polarization micrograph of a transverse section of the grown boule. Left part is a single domain; right part is heavily multi-domain. (b) Sketch of the unstable molten zone, which is likely to drop down. (c) Optical polarization micrograph of a solid-liquid interface. (d) Enlarged solid-liquid interface of the single domain part in (c). (e) Enlarged solid-liquid interface of the multi domain part in (c).
\par
\noindent
\\
Fig. 8.
Crystal growth of Bi-2223. (a) Premelted feed rod at 25 mm/h. (b) Actual crystal growth images near molten zone, taken once every week. (c) Obtained Bi-2223 boule. Large crystals were found at the last 1-cm part. (d) Optical poralization micrograph of a transverse section of the last 1-cm part of the boule. (e) Grown crystals cleaved from the upper right side of (d).
\par
\noindent
\\
Fig. 9.
DC susceptibility of the as-grown Bi-2223 single crystal.
\par
\noindent
\\
Fig. 10.
Equilibrium oxygen partial pressures $P_{O2}$ versus reciprocal temperature for a Bi$_{2}$Sr$_{2}$CaCu$_2$O$_{8+\delta}$ single crystal for various oxygen contents ($\delta$). Triangles, squares, circles, and crosses are experimentally obtained data for $\delta = 0.29$, $\delta = 0.28$, $\delta = 0.27$, and $\delta = 0.26$, respectively.
\par
\noindent
\\
Fig. 11.
DC susceptibilities of the Bi$_{2}$Sr$_{2}$CaCu$_2$O$_{8+\delta}$ single crystals for $\delta = 0.22$, $\delta = 0.24$, $\delta = 0.26$, and as-grown. Data were normalized at the lowest temperature.
\par
\noindent
\\
Fig. 12.
Oxygen content dependence of $T_c$ in Bi-2212.
\par
\noindent
\\
Fig. 13.
Furnace for the precise oxygen control.
\par
\noindent
\\
Fig. 14.
In-plane resistivities $\rho_a$ of Bi$_{2}$Sr$_{2}$CaCu$_2$O$_{8+\delta}$ single crystals versus temperature for various oxygen contents ($\delta$). The solid straight lines, which are linear extraporation of $\rho_a$ at higher temperatures, are eye guides for the near optimally doped ($\delta = 0.24$) and underdoped ($\delta$ = 0.22, 0.217, and 0.2135) samples. The temperatures $T^\ast_{\rho_a}$ at which the $\rho_a$ deviates from $T$-linear behavior are shown by arrows for the underdoped samples.
\par
\noindent
\\
Fig. 15.
(a) Out-of-plane resistivities $\rho_c(T)$ of a Bi$_{2}$Sr$_{2}$CaCu$_2$O$_{8+\delta}$ single crystal (sample I) for various oxygen contents (0.2135 $\le$ $\delta$ $\le$ 0.28). (b) Out-of-plane resistivity $\rho_{c}(T)$ of a Bi$_{2}$Sr$_{2}$CaCu$_2$O$_{8+\delta}$ single crystal (sample II) for oxygen content 0.26$\le$$\delta$$\le$0.3. The solid straight lines, which are linear extraporations of $\rho_{c}$ at higher temperatures, are eye guides for the overdoped ($\delta$=0.26, 0.27, and 0.28) samples. Arrows indicate the temperatures $T^\ast_{\rho_c}$ below which $\rho_{c}$ shows a typical upturn.
\par
\noindent
\\
Fig. 16.
Magnetic susceptibilities (a) ${\chi}_{c}(T)$ $(H\parallel{c})$ and (b) ${\chi}_{ab}(T)$ $(H\perp{c})$ of the Bi$_{2}$Sr$_{2}$CaCu$_2$O$_{8+\delta}$ single crystals with various oxygen contents ($\delta$) under H=5 T. The solid straight lines in (a), which are linear extraporations of $\chi_c$ at higher temperatures, are eye guides for the overdoped ($\delta$=0.26, 0.27, and 0.28) samples. Arrows indicate the temperatures $T^\ast_{\chi}$ at which $\chi_c$ starts to decrease from its linear high-temperature behavior.
\par
\noindent
\\
Fig. 17.
Scaling behavior for the resistivity ratio $r = \rho_{c,peak}/\rho_{c}(RT)$ versus doping levels ($p$).
\par
\noindent
\\
Fig. 18.
Characteristic temperatures obtained by various measurements versus carrier concentration $p$. The $p$ values (including for the previous results) were systematically estimated from the empirical $T_c$ vs. $p$ relation in the text.
\par
\noindent
\\
Fig. 19.
$\chi_c(T)$ versus $\chi_{ab}(T)$ plots for (a) underdoped ($\delta$=0.22), (b) slightly overdoped ($\delta$=0.26) and (c) heavily overdoped ($\delta$$\approx$0.3) Bi$_{2}$Sr$_{2}$CaCu$_2$O$_{8+\delta}$ single crystals. Several typical temperatures are shown by arrows.
\par
\noindent
\\
Fig. 20.
Doping dependence of the tunneling DOS in the normal state.
\par
\noindent
\\
Fig. 21.
Isotropic component of the spin susceptibilities, $\chi^{iso}_{spin}(T)$, of a Bi$_{2}$Sr$_{2}$CaCu$_2$O$_{8+\delta}$ single crystal for oxygen content $\delta$=0.22 (underdoped) and 0.25 (optimum doped). The solid lines are numerical fits, assuming the BCS-type DOS for the pseudogap.
\par
\noindent
\\
Fig. 22.
(a) Normalized transition temperature $T_c$ plotted against the $c$-axis variation. The $T_c$ and $c$-axis length of optimum-doping samples are 89 K, 30.864 $\AA$ for Bi-2212 and 108 K, 37.119 $\AA$ for Bi-2223, respectively. (b) Thermopower $S$ of Bi$_2$Sr$_2$Ca$_2$Cu$_3$O$_{10+\delta}$ single crystal annealed in various atmospheres.
\par
\noindent
\\
Fig. 23.
In-plane resistivity $\rho_{ab}$ of Bi$_2$Sr$_2$Ca$_2$Cu$_3$O$_{10+\delta}$ single crystal annealed in various atmospheres. The solid straight lines, which are linear extrapolations of $\rho_{ab}$ at higher temperatures, are eye guides. The temperatures $T^*_{\rho_{ab}}$ at which the $\rho_{ab}$ deviates from $T$-linear behavior are shown by arrows. The scale is expanded in the inset for a better view around $T_c$.
\par
\noindent
\\
Fig. 24.
Out-of-plane resistivity $\rho_c$ of Bi$_2$Sr$_2$Ca$_2$Cu$_3$O$_{10+\delta}$ single crystal annealed in various atmospheres. Overdoped behavior of out-of-plane resistivity is shown in the inset. The solid straight lines in the inset, which are linear extrapolations of $\rho_c$ at higher temperatures are eye guides for the overdoped sample. Arrows indicate the temperatures $T^*_{\rho_c}$ below which $\rho_c$ shows a characteristic upturn.
\par
\noindent
\\
Fig. 25.
Crystal structure of Bi$_2$Sr$_2$Ca$_2$Cu$_3$O$_{10+\delta}$. Inner square-planar CuO$_2$ plane and outer pyramidal CuO$_2$ planes are shown schematically. Bold lines denote a unit cell.
\par
\noindent
\\
Fig. 26.
Schematic phase diagram of high-$T_c$ superconductors (for summary).
\par
\noindent
\\

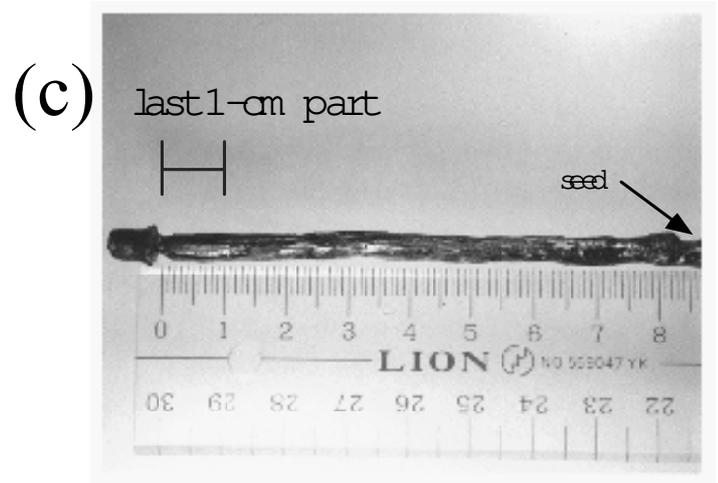
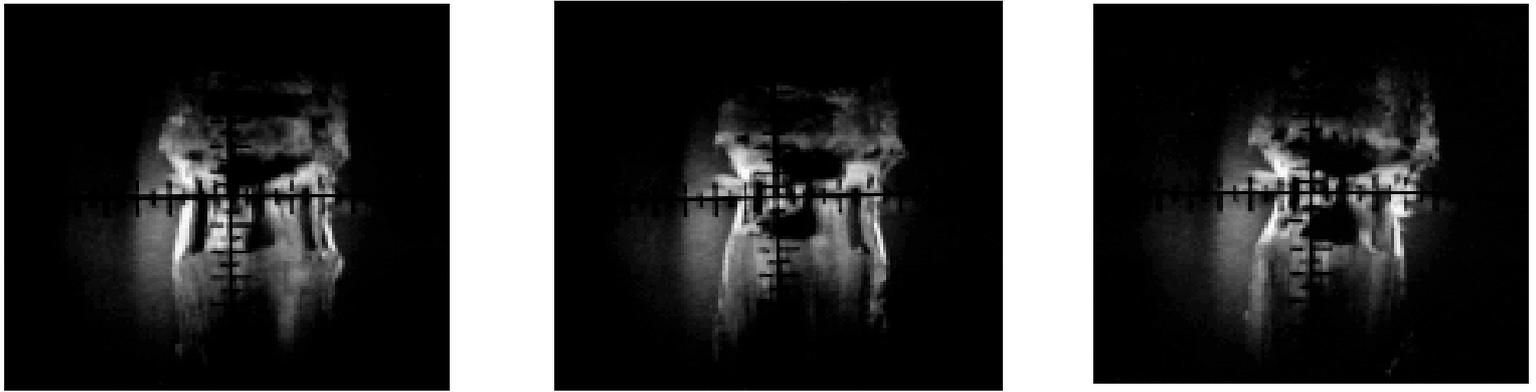
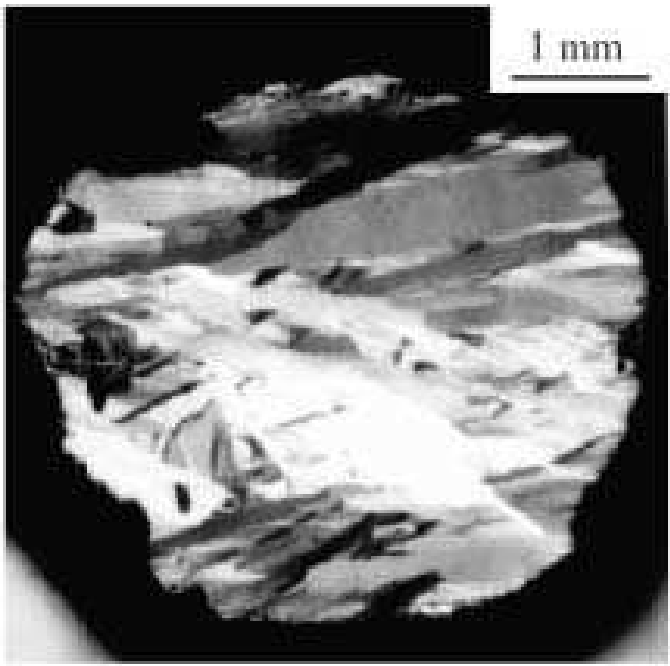
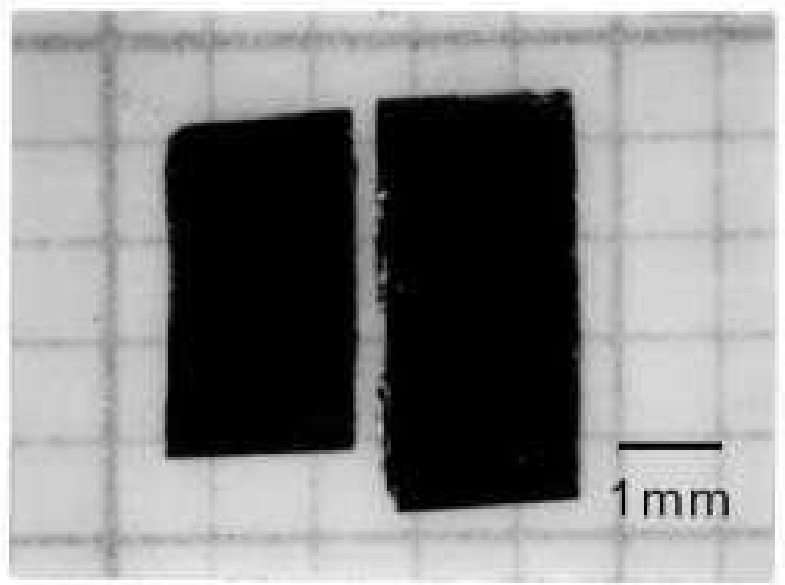
\end{document}